\documentclass[12pt]{article}
\usepackage{amssymb}

\newcommand{\beqn}{\begin{equation}}
\newcommand{\eeqn}{\end{equation}}
\newcommand{\bear}{\begin{eqnarray}}
\newcommand{\eear}{\end{eqnarray}}
\newcommand{\bean}{\begin{eqnarray*}}
\newcommand{\eean}{\end{eqnarray*}}

\begin{document}
\begin{center} \Large  On some mathematical model of turbulent flow with intensive selfmixing
  \end{center}

\vspace{0.1cm}

\begin{center}
M. BURNAT\\
\vspace{0.3cm}

{\bf{University of Warsaw\\
Institute of Applied Mathematics and Mechanics}}\\
Banacha 2, 02-097 Warszaw, Poland
\end{center}

\vspace{0.5cm}

The paper presents a new possibility of mathematical modeling of turbulent flow with the intensive selfmixing. The flow is described in this model by a family of mappings
\[
\Phi_{t,t^o}: \omega \to S(t,t^o,\omega); \qquad \omega,S \subset R^3, t > t^o
\]
such that in general
\[
\omega^1 \cap \omega^2 = \emptyset \Rightarrow\!\!\!\!\!\!/ \;\;S(t,t^o,\omega^1)\cap S(t,t^o,\omega) = \emptyset
\]

The model allows a new approach to certain problems of turbulent flow. For example there are possibilities to introduce some models of turbulent chemical reactors, or description of the unpredictable explosions of turbulence in the calm laminate flow. In this model we introduce mathematical tools that allow definition of various mixing states of the fluid. It is possible to formulate in the appropriate mathematical language, the idea that the given fluid can have various mixing states. It is also possible to determine the influence of a given mixing state on the flow parameters such as velocity field and density. The first formulation of the model is given in [1].

\vspace{0.5cm}
{\large{1.{\underline{The general feature of the model}}}}
\vspace{0.5cm}

For the balls
\[
\omega_\varepsilon (x) = \{y \in R^3:|y-x|<\varepsilon\},  \varepsilon >0,
\]
consider the sets ${\bf{V}}_\varepsilon(x,t)\subset R^3$ of velocities of fluid particles moving at the time $t$ in $\omega_\varepsilon(x)$. The velocity sets may be measured. The laminar model and our model may be shortly characterized in the following way.\\

\underline{Laminar model}\\

1) The flow is described by family of regular invertible mappings (diffeomorphisms):
\[
\Phi_{t,t^o}: \omega \Longleftrightarrow S(t,t^o,\omega);  \qquad \omega  S(t,t^o,\omega) \subset R^3, \qquad t > t^o,
\]
where $S(t,t^o,\omega)$ is the set containing at $t>t^o$ the whole fluid contained at $t^o$ in $\omega$, and no another one. It means that \underline{the fluid contained in $\omega$ at the time $t$ does not mix with}
\underline{any fluid} in the neighbourhood, and \underline{keep for ever its identity}.\\

2) There exist the Euler velocity field $v(x,t)$,\\
\\
(1) $\hspace{5cm} v(x,t) = \lim_{\varepsilon \to 0} {\bf{V}}_\varepsilon(x,t).$\\

\underline{Our model}.\\

I. The mappings
\[
\Phi_{t,t^o}: \omega \to S(t,t^o,\omega)
\]
describing the flow are not invertible. The set $S(t,t^o,\omega)$ \underline{is the minimal set containing} \underline{at the time $t >t^o$ the whole fluid $m(\omega,t^o)$}, but it may contain another fluid portions.\\

II. Instead of the field of Euler velocities (1) we assume only the existence of field of velocity sets
\[
{\bf{V}}(x,t) \subset R^3,
\]
such that for numbers $\Delta>0, \varepsilon >0$ small enough the following approximation\\
\\
(2) $\hspace{4cm} S(t+\Delta,t,\omega_\varepsilon(x)) \approx x + \Delta {\bf{V}}(x,t)$\\
\\
holds in the sense of measure.

Basing on the assumptions I, II we can formulate the integral conservation laws of mass, impulse, energy and impulse momentum. Moreover, we obtain the equivalent closed integro-differential system. In this way we obtain a model of fluid flow in which different fluid portions are mixing one with another loosing their identity.

To this end we have to introduce first the notion of \underline{the physical $\alpha$-quantities}. We begin with the definition of the \underline{$\alpha$-density}. This is a nonnegative function
\[
\varrho:R^7 \to R, \qquad \varrho = \varrho(x,t,\alpha) = \left\{
\begin{array}{ll}
> 0, & \alpha \in {\bf{V}}(x,t),\\
\quad 0, & \alpha \in R^3 \backslash {\bf{V}}(x,t)
\end{array}
\right.
\]
with the following properties. For the mass portion filling up $\omega$ at the time $t$, the following equalities hold
\[
|m(\omega,t)| = \int_\omega \varrho(x,t)dx = \kappa \int_\omega \int_{{\bf{V}}(x,t)} \varrho(x,t,\alpha)dx d\alpha,
\]
\[
\varrho(x,t) = \kappa \int_{{\bf{V}}(x,t)}\varrho(x,t,\alpha)d\alpha
\]
where $\varrho(x,t)$ is the usual density, and $\kappa>0$  is some scaling constant.

We shall construct an approximation of $\varrho(x,t,\alpha)$ explaining its physical sense.\\
The sets
\[
S(t+\Delta,t,\omega_\varepsilon(x)) = \Phi_{t+\Delta,t}(\omega),
\]
where $\Delta>0,\varepsilon >0$ small enough, are the minimal sets that contain at the time $t+\Delta$ the whole fluid filling up $\omega_\varepsilon(x)$ at the time $t$. This fluid has in $S(t+\Delta,t,\omega_\varepsilon(x))$ at $t+\Delta$ density
\[
g(y)\;\mbox{for}\; y \in S(t+\Delta,t,\omega_\varepsilon(x).
\]
Using (2), we have
\[
|m(\omega_\varepsilon(x),t)|= \int_{S(t+\Delta,t,\omega_\varepsilon(x)}\!\!\!\!\!\!\!\!\!\!g(y)dy \approx \int_{x+\Delta{\bf{V}}(x,t)}\!\!\!\!\!\!\!\!\!\!g(y)dy = \Delta^3 \int_{{\bf{V}}(x,t)}\!\!\!\!\!\!\!\!g(x+\alpha\Delta)d\alpha.
\]
Hence
\[
\varrho(x,t) \approx \frac{|m(\omega_\varepsilon(x),t)|}{\frac{4}{3}\pi\varepsilon^3} \approx \frac{3}{4\pi}\biggl(\frac{\Delta}{\varepsilon}\biggl)^3 \int_{{\bf{V}}(x,t)}\!\!\!\!g(x+\alpha\Delta)d\alpha,
\]
and we obtain the following approximation of the $\alpha$-density
\[
\varrho(x,t,\alpha) \approx \left\{
\begin{array}{ll}
g(x+\Delta\alpha), & \alpha \in {\bf{V}}(x,t)\\
\qquad 0, & \alpha \in R^3 \backslash {\bf{V}}(x,t)
\end{array}
\right.
,\mbox{and}\; \kappa = \frac{3}{4\pi}\biggl(\frac{\Delta}{\varepsilon}\biggl)^3.
\]
In our model we use the general notion of the \underline {$\alpha$ - quantity}. For example we define the \underline
{$\alpha$ - impulse} as $\alpha \varrho(x,t,\alpha)$, so that the impulse of the fluid portion $m(\omega,t)$ is equal to
\[
imp(\omega,t) = \kappa \int_\omega \int_{{\bf{V}}(x,t)} \alpha \varrho(x,t,\alpha)dx d\alpha 
\]
If some physical expression contains the k-fold integration over the variable $\alpha$, then we multiply the integral by $\kappa^k$.

Given the $\alpha$-quantities we may obtain the observed mean Euler velocity $v(x,t)$ of the flow:
\[
v(x,t) = \lim_{\stackrel{d(\omega)\to 0}{x \in \omega}} \frac{imp(\omega,t)}{|m(\omega,t)|} = \frac{\int_{R^3}\alpha\varrho(x,t,\alpha)d\alpha}{\int_{R^3}\varrho(x,t,\alpha)d\alpha}
\]
In almost all physical situations one may take some ball $A \subset R^3$, for which we know that $\varrho(x,t,\alpha) = 0$ for $\alpha \in R^3 \backslash A$, and we may integrate over A only.

\vspace{0.5cm}
{\large{2.{\underline{The Mass Conservation Law}}}}
\vspace{0.5cm}

We have to introduce the fluid portions $m(\omega,a,t), a \subset A$, which are parts of the fluid in $\omega$ moving at the time $t$ with the velocities $\alpha \in a \subset A$. We have
\[
|m(\omega,a,t)| = \kappa \int_\omega \int_a \varrho(x,t,\alpha)dx d\alpha,  \qquad |m(\omega,A,t)| = |m(\omega,t)|.
\]
Moreover, we introduce the \underline{mass mixer}:
\[
M: R^{10} \to R, \qquad M = M(x,t,\alpha,\beta),
\]
which describes the following mixing processes:
\[
\underline{m(\omega,a,t)}_{\;\;\stackrel{\Longleftrightarrow}{a,b \subset A}\;\;}  m(\omega,b,t) \qquad
\begin{tabular}{|c|} \hline
\\
$\kappa^2 \int_\omega \int_a \int_{b\backslash a}\!\!\! M(x,t,\alpha,\beta)dx d\alpha d\beta$
\\
\\ \hline
\end{tabular}
\]
(3)
\[
m(\omega,a,t)_{\;\;\stackrel{\Longleftrightarrow}{a,b \subset A}\;\;} \underline{m(\omega,b,t)} \qquad
\begin{tabular}{|c|} \hline
\\
$\kappa^2 \int_\omega \int_b \int_{a\backslash b}M(x,t,\alpha,\beta)dx d\alpha d\beta$ 
\\
\\ \hline
\end{tabular}
\]
\\
If in the mixing process one of the portions $m(\omega,a,t), m(\omega,b,t)$ is underlined, then it means that we ask  amount of mass (positive or negative equal to the integrals in the frames on the right) which is transported to the underlined portion in the unit time . It turns out that
\[
M(x,t,\alpha,\beta) = -M(x,t,\beta,\alpha)
\]
The following three mixing processes (4), (5), (6) should be taken into account in the mass conservation low for the portion $m(\omega,a,t)$, where $a \subset A$:\\
\\
(4) $\hspace{4cm} \underline{m(\omega,a,t)} \Longleftrightarrow m(\omega,A \backslash a,t)$
\\
\\
(5) $\hspace{4cm} \underline{m(\omega,a,t)} \Longleftrightarrow m(R^3 \backslash \omega,a,t)$
\\
\\
(6) $\hspace{4cm} \underline{m(\omega,a,t)} \Longleftrightarrow m(R^3 \backslash \omega,A \backslash a,t).$
\\

The amount of mass transported in the unit time in this processes to the portion $m(\omega,a,t)$ is respectively equal to\\
\\
(4a) $\hspace{4cm}$
\begin{tabular}{|c|} \hline
\\
$\kappa^2 \int_\omega \int_a \int_{A\backslash a}\!\!\! M(x,t,\alpha,\beta)dx d\alpha d\beta$ 
\\
\\ \hline
\end{tabular}
\\
\vspace{0.2cm}
\\
(5a) $\hspace{1cm}$
\begin{tabular}{|c|} \hline
\\
$-\kappa \int_{\partial\omega} \int_a [<\alpha,n(x)\varrho(x,t,\alpha)>-E<n(x), \mbox{grad}_x \varrho(x,t,\alpha)>]dx d\alpha$ 
\\
\\ \hline
\end{tabular}
\\
\vspace{0.2cm}
\\
(6a) $\hspace{2.5cm}$
\begin{tabular}{|c|} \hline
\\
$-\kappa^2 \int_{\partial\omega} \int_a \int_{A\backslash a} < n(x), B(x,t,\alpha,\beta) > dx d\alpha d\beta$
\\
\\ \hline
\end{tabular}
\\
\\
where $n(x)$ is the external normal unit vector to the boundary $\partial\omega$.

In (4) we used the properties (3) of the mass mixer $M(x,t,\alpha,\beta)$. We consider the process (5) like in the laminar model without any mixer, so as in this process we have on the both sides of the boundary $\partial\omega$, fluids with the same kinematical characteristics. In the process (5) we take into account the mass transport of the diffusion type with the constant $E>0$. The process (6) is a mixing process of two fluid portions with different kinematical characteristics across the boundary $\partial\omega$. Therefore we have to introduce \underline{the boundary mass mixer B},
\[
B: R^{10} \to R^3, \qquad \qquad B = B(x,t,\alpha,\beta).
\]
It describes the amount of mass transported in the unit time to $m(\omega,a,t)$ in the process (6).

Finally taking together (4a), (5a), (6a) we obtain the following integral mass conservation law for the portion $m(\omega,a,t)$.\\
\\
(7) $\hspace{2cm}$
\begin{tabular}{|c|} \hline
\\
$\int_\omega \int_a \biggl\{\partial_t \varrho(x,t,\alpha) + <\alpha, \mbox{grad}_x \varrho> -E \Delta_x \varrho(x,t,\alpha) + $\\
$\kappa \int_{A \backslash a} \biggl[\mbox{div}_x B(x,t,\alpha,\beta) - M(x,t,\alpha,\beta)\biggl]d\beta \biggl\}dx d\alpha=0$\\
$\hspace{7cm} \omega \subset R^3, a \subset A$\\
\\ \hline
\end{tabular}
\\
\\
We shall use the following simple\\

\underline{Localization Theorem}\\

{\it{
Let $D: R^7 \to R, F: R^7 \to R$. Then the condition\\
\\
(8) $\hspace{3cm} \int_\omega \int_a \biggl[ D(x,t,\alpha)+ \kappa \int_{A\backslash a} F(x,t,\alpha,\beta)d\beta  \biggl]dx d\alpha = 0,$
\[
\hspace{8cm} \omega \subset R^3,a \subset A
\] 
is equivalent to the following system of equations:\\
\\
(9)$\hspace{4cm} D(x,t,\alpha)+ \kappa \int_A F(x,t,\alpha,\beta)d\beta = 0,$
\[
\hspace{6cm} x\in R^3,\alpha \in A
\]
\\
(10) $\hspace{4cm} F(x,t,\alpha,\beta) + F(x,t,\beta,\alpha) = 0,$
\[
\hspace{6cm} x \in R^4, \alpha,\beta \in A
\]
}}

Applying the Localization Theorem to integral conservation law (7), we obtain the following integro-differential system equivalent to the integral law:\\
\\
(11) 
\begin{tabular}{|c|} \hline
\\
$\partial_t\varrho(x,t,\alpha)+<\alpha,\mbox{grad}_x\varrho>-E\Delta_x\varrho+\kappa\int_A\biggl[\mbox{div}_xB(x,t,\alpha,\beta)-M(x,t,\alpha,\beta)\biggl]d\beta=0$\\
$\hspace{12.5cm} x\in R^3,\alpha\in A$\\
$\mbox{div}_xB(x,t,\alpha,\beta)-M(x,t,\alpha,\beta)+\mbox{div}_xB(x,t,\beta,\alpha)-M(x,t,\beta,\alpha)=0$\\
$\hspace{12cm} x\in R^3,\alpha,\beta\in A$\\
\\ \hline
\end{tabular}

\vspace{0.5cm}
{\large{3.{\underline{The Impulse Conservation Law}}}}
\vspace{0.5cm}

We introduce the following impulse mixer $J(x,t,\alpha,\beta)$. It is the function
\[
J:R^{10} \to R^3, \quad J(x,t,\alpha,\beta) = \alpha M(x,t,\alpha,\beta)+i(x,t,\alpha,\beta)
\]
where $\alpha M(x,t,\alpha,\beta)$ describes the transport of impulse caused by the mass transport, and $i(x,t,\alpha,\beta)$ describes other types of impulse changes. The physical sense of the mixer $J(x,t,\alpha,\beta)$ is the following. The mixer $J(x,t,\alpha,\beta)$ describes the amount of impulse transported in a unit time to the portion $m(\omega,a,t)$ in the process
\[
\underline{m(\omega,a,t)}_{\;\;\stackrel{\Longleftrightarrow}{a,b\subset A}\;\;} m(\omega,b,t).
\]
This amount is equal to\\
\\
(12) $\hspace{4cm} \kappa \int_\omega \int_a \int_{b\backslash a}J(x,t,\alpha,\beta)dx d\alpha d\beta.$
\\

On the one hand, the change of the impulse in a unit time in the portion $m(\omega,a,t)$ is equal to
\[
\kappa\partial_t \int_\omega \int_a \alpha\varrho(x,t,\alpha)dx d\alpha.
\]

On the other hand, this change is equal to the force acting on $m(\omega,a,t)$. This force is the sum of the following forces\\
\\
\begin{tabular}{|c|} \hline
\\
The forces acting on $m(\omega,a,t)$ by fluid portions:\\
$m(\omega,A\backslash a,t), m(R^3\backslash \omega,a,t), m(R^3\backslash \omega,A \backslash a,t)$.
\\
\\ \hline
\end{tabular}
+
\begin{tabular}{|c|} \hline
\\
The external forces acting\\
on $m(\omega,a,t).$
\\
\\\hline
\end{tabular}
\\

Let us determine this forces. The forces acting on $m(\omega,a,t)$ are equal to the amount of impulse transported in a unit time to $m(\omega,a,t)$.

In the process\\
\[
\underline{m(\omega,a,t)} \Longleftrightarrow m(\omega,A \backslash a,t)
\]
(according to the definition of the impulse mixer) the amount of impulse transported to $m(\omega,a,t)$ in a unit time is equal to
\[
\kappa^2\int_\omega \int_a \int_{A \backslash a} \!\!\! J(x,t,\alpha,\beta)dx d\alpha d\beta.
\]

In the process\\
\[
\underline{m(\omega,a,t)} \Longleftrightarrow m(R^3 \backslash \omega,a,t),
\]
the amount of impulse transported in a unit time to $m(\omega,a,t)$, is equal to:
\[
-\kappa \int_{\partial\omega}\int_a \alpha \biggl[<\alpha,n(x)>\varrho(x,t,\alpha)-E<n(x),grad_x\varrho>\biggl]dx d\alpha
\]

In the process\\
\[
\underline{m(\omega,a,t)} \Longleftrightarrow m(R^3 \backslash a,A \backslash a,t)
\]
the increase of impulse in the unit time in the portion $m(\omega,a,t)$ is equal to
\[
-\kappa^2 \int_{\partial \omega}\int_a \int_{A\backslash a}\!\!\!\! J_B(x,t,\alpha,\beta)\cdot n(x)dx d\alpha d\beta
\]
where we introduced the boundary impulse mixer $J_B(x,t,\alpha,\beta)$. It is a matrix $J_B:R^{10} \to R^3\times R^3$. Finally, let us take into account the processes (4),(5),(6), and the external forces. Then, after changing the surface integrals into volume ones, we obtain the following integral form of the impulse conservation law for the portion $m(\omega,a,t)$:\\
\[
\int_\omega \int_a \biggl\{ \partial_t\alpha\varrho(x,t,\alpha)+ \alpha\biggl[ <\alpha,grad_x \varrho>-E\Delta_x\varrho\biggl]+
\]
(13) $\hspace{1cm} +\kappa\int_{A\backslash a}[\mbox{div}_xJ_B(x,t,\alpha,\beta)-J(x,t,\alpha,\beta)]d\beta - f(x,t,\alpha)\biggl\}dx d\alpha = 0,$
\[
\hspace{9cm}\omega \subset R^3, a \subset A
\]
where
\[
F(t) = \kappa \int_\omega \int_a f(x,t,\alpha)dx d\alpha
\]
is the external force acting on $m(\omega,a,t)$. In the case of the earth gravitation field, we have
\[
f(x,t,\alpha) = g\varrho(x,t,\alpha),  \quad g = \mbox{const}.
\]
Applying to (13) the Localization Theorem, we obtain the following integro-differential system equivalent to the integral impulse conservation law:\\
\\
(14) $\hspace{2cm}$
\begin{tabular}{|c|} \hline
\\
$\partial_t\alpha\varrho(x,t,\alpha)+\alpha\ [ <\alpha,grad_x \varrho> - E\Delta_x\varrho] =$\\
$\hspace{7cm} x \in R^3, \alpha \in A$\\
$= \kappa \int_A[J(x,t,\alpha,\beta)-\mbox{div}_x J_B(x,t,\alpha,\beta)]d_\beta + f(x,t,\alpha)$\\
\\
$J(x,t,\alpha,\beta)+J(x,t,\beta,\alpha) - \mbox{div}_x[J_B(x,t,\alpha,\beta)+J_B(x,t,\beta,\alpha)] = 0$\\
$\hspace{9cm} x \in R^3; \alpha,\beta \in A$\\
\\ \hline
\end{tabular}

\vspace{0.5cm}
{\large{4.{\underline{The Energy Conservation Law.}}}}
\vspace{0.5cm}

Let us introduce the $\alpha$-inner energy of the flow. This is a function
\[
\varepsilon :R^7 \to R, \varepsilon = \varepsilon(x,t,\alpha),
\]
\[
\varepsilon(x,t.\alpha) = 0 \; \mbox{for}\; \alpha \in A \backslash V(x,t).
\]
Denote by $e(\omega,a,t)$ the inner energy of the fluid portion $m(\omega,a,t)$. Then the physical sense of the $\alpha$-energy is explained by the formula:
\[
e(\omega,a,t) = \kappa \int_\omega \int_a \varepsilon(x,t,\alpha) \varrho(x,t,\alpha)dx d\alpha.
\]
Denoting by $e(\omega,t)$ the inner energy of the fluid portion $m(\omega,t)$, we obtain the following equality:
\[
e(\omega,t) = \int_\omega \varepsilon(x,t)\varrho(x,t)dx = e(\omega,A,t) = \kappa \int_\omega \int_A e(x,t,\alpha)\varrho(x,t,\alpha)dx d\alpha.
\]
Hence, we get\\
\\
(15) $\hspace{3cm} \varepsilon(x,t)\varrho(x,t) = \kappa \int_A \varepsilon(x,t,\alpha)\varrho(x,t,\alpha)d\alpha.$
\\

The force $F(\omega,a,t,)$ acting on the portion $m(\omega,a,t)$ is equal to the increase of the impulse of $m(\omega,a,t)$ in the unit time. Hence, from the impulse conservation law we obtain
\[
F(\omega,a,t) = \kappa \int_\omega \int_a E(x,t,\alpha)dx d\alpha,
\]
where
\[
E(x,t,\alpha) = \kappa\int_{A \backslash a}\!\![J(x,t,\alpha) - \mbox{div}_x J_B(x,t,\alpha,\beta)]d\beta + \alpha[-<\alpha, grad_x \varrho> + E\Delta_x \varrho] + f(x,t,\alpha).
\]

The increase of the energy of the portion $m(\omega,a,t)$ in the unit time is equal to the power due to the force $F(\omega,a,t)$. Let us evaluate it. Divide $\omega$ and $a$ into small parts $\Delta \omega \subset \omega, \Delta a \subset a$, and notice that $F(\omega,a,t)$ acts on $m(\omega,a,t)$ is such a way that on the small parts $m(\Delta\omega, \Delta a, t) \subset m(\omega,a,t)$ the following forces are acting:
\[
F(\Delta\omega,\Delta a,t) = \kappa \int_{\Delta\omega}\int_{\Delta a}\!\!\!E(x,t,\alpha)dx d\alpha.
\]
The power due to $F(\omega,a,t)$ is the sum of the powers $P(\Delta\omega,\Delta a,t)$ due to the forces $F(\Delta \omega,\Delta a,t)$ where $\Delta\omega \subset \omega, \Delta a \subset a$. For $\stackrel{o}{x} \in \Delta \omega, \stackrel{o}{\alpha} \in \Delta a$, we obtain the following approximation for small $\Delta t$: 
\[
P(\Delta\omega,\Delta a,t) \approx \frac{1}{\Delta t}\biggl(|\stackrel{o}{\alpha}|\Delta t\biggl)\biggl(\kappa |\Delta \omega||\Delta a|\biggl)|E(\stackrel{o}{x},t,\stackrel{o}{\alpha})|\cos \nless(\stackrel{o}{\alpha},E(\stackrel{o}{x},t,\stackrel{o}{\alpha})) =
\]
\[
= \kappa |\Delta \omega ||\Delta a| < \stackrel{o}{\alpha}, E(\stackrel{o}{x},t,\stackrel{o}{\alpha})>
\]
Finally, for the whole power $P(\omega,a,t)$ due to the  force $F(\omega,a,t)$, we obtain
\[
P(\omega,a,t) = \sum_{\Delta\omega \subset a,\Delta a \subset a}\!\!\! P(\Delta \omega,\Delta a,t) = \kappa \int_\omega \int_a \!< \alpha, E(x,t,\alpha )> dx d\alpha.
\]
Now, we can write down the integral energy conservation law for the portion $m(\omega,a,t)$:\\
\\
(16) $\hspace{0.5cm} \int_\omega \int_a \biggl\{ \partial_t \biggl[\varepsilon(x,t,\alpha)\varrho(x,t,\alpha)+|\alpha|^2 \frac{\varrho(x,t,\alpha)}{2}\biggl]-<\alpha,E(x,t,\alpha)>\biggl\}dx d\alpha = 0,$
\[
\omega \subset R^3, a \subset A
\]
Putting $a=A$ and taking into account (15), we obtain the following relation\\
\\
(17) $\hspace{0.3cm} \partial_t(\varepsilon(x,t)\varrho(x,t)) = \kappa \int_A \biggl[<\alpha,f>-|\alpha|^2 \biggl(<\alpha,grad_x \varrho>-E \Delta_x \varrho + \partial_t \frac{\varrho(x,t\alpha)}{2}\biggl)\biggl]d\alpha$.
\\

In this way, one determine the mean inner energy $\varepsilon(x,t)$ for given $\alpha$-density $\varrho(x,t,\alpha).$ 

Applying to (16) the Localization Theorem, we obtain the following integro-differential system equivalent to the integral energy conservation law:\\
\\
(18) $\hspace{0.5cm} $ 
\begin{tabular}{|c|} \hline
\\
$\partial_t\biggl[ \varepsilon(x,t,\alpha)\varrho(x,t,\alpha)+|\alpha|^2 \frac{\varrho}{2}\biggl] + |\alpha|^2 \biggl[<\alpha,grad_x \varrho>-E\Delta_x \varrho \biggl]=$\\
$\hspace{10cm}x\in R^3,\alpha \in A$\\
$= \int_A\biggl[<\alpha,J(x,t,\alpha,\beta)-\mbox{div}_xJ_B(x,t,\alpha,\beta)>\biggl]d\beta + <\alpha,f(x,t,\alpha>$\\
\\
$<\alpha,J(x,t,\alpha,\beta)- \mbox{div}_xJ_B(x,t, \alpha,\beta)> +$\\
\\
$+ <\beta,J(x,t,\beta,\alpha)- \mbox{div}_x J_B(x,t,\beta,\alpha)>=0$\\
$\hspace{7cm}x\in R^3, \alpha,\beta \in A$ \\
\\ \hline
\end{tabular}

\vspace{0.5cm}
{\large{5.{\underline{The Impulse Momentum Conservation Law.}}}}
\vspace{0.5cm}

Applying the principle\\
\\
\begin{tabular}{|c|}\hline 
\\
The derivative $\partial_t$of the impulse momentum
\\
\\ \hline
\end{tabular}
=
\begin{tabular}{|c|} \hline
\\
The momentum of acting forces
\\
\\\hline
\end{tabular}\\
\\
\\
and our Impulse Conservation Law we can formulate the \underline{Impulse Momentum Conservation} \underline{Law} for the fluid portion $m(\omega,a,t)$. \\

If we denote $F \wedge H = ((F \wedge H)_1, (F \wedge H)_2, (F \wedge H)_3)$, for $F,H \in R^3$, then we obtain the following integro-differential system equivalent to the integral Impulse Momentum Conservation Law\\
\\
(19) $\hspace{0.3cm} $ 
\begin{tabular}{|c|} \hline
\\
$\partial_t(x \wedge \alpha\varrho)_j= \mbox{div}_x \biggl[E(x\wedge \alpha)_j grad_x \varrho - \alpha(x \wedge \alpha)_j\biggl] - (x\wedge f(x,t,\alpha))_j +$\\
$+ \kappa \int_A\biggl[\biggl(x \wedge J(x,t,\alpha,\beta)\biggl)_j + \partial_{x_1}(x \wedge h^1)_j + \partial_{x_2}(x \wedge h^2)_j + \partial_{x_3}(x \wedge h^3)_j\biggl]d\beta$\\
$j = 1,2,3, x \in R^3, \alpha \in A$\\
$\biggl(x \wedge J(x,t,\alpha,\beta)\biggl)_j + \partial_{x_1}\biggl(x \wedge h^1(x,t,\alpha,\beta)\biggl)_j + \partial_{x_2}(x \wedge h^2)_j + \partial_{x_3}(x \wedge h^3)_j + $\\
$+\biggl(x \wedge J(x,t,\beta,\alpha)_j +\partial_{x_1}\biggl(x \wedge h^1(x,t,\beta,\alpha)\biggl)_j + \partial_{x_2}(x \wedge h^2)_j + \partial_{x_3}(x \wedge h^3)_j = 0$\\
$\hspace{9cm}j = 1,2,3, x \in R^3, \alpha,\beta \in A$ \\
\\ \hline
\end{tabular}\\ 
\\
where the matrix $J_B$ is denoted as
\[
J_B(x,t,\alpha,\beta) = (h^i_j),h^i(x,t,\alpha,\beta) = (h^i_1, h^i_2, h^i_3) 
\] 

\vspace{0.5cm}
{\large{6.{\underline{The constitutive relation and the full closed system of the model.}}}}
\vspace{0.5cm}

In order to formulate the constitutive relation let us express approximately  the boundary mixer $B(x,t,\alpha,\beta)$ as a function of the mixer $M(x,t,\alpha,\beta)$. To this end, introduce the following notations
\[
\partial^\varepsilon \omega = \bigcup_{x \in \partial\omega} \biggl\{y :|y-x|< \varepsilon \biggl\}
\]
\[
D^\varepsilon \omega = \partial^\varepsilon \omega \backslash \omega \subset R^3 \backslash \omega, \quad d^\varepsilon \omega = \partial^\varepsilon \omega \cap \omega \subset \omega,
\]
where $\varepsilon$ is a positive small number.

Consider two following complex mixing processes $(1^o)$ and $(2^o)$. These processes , taken together, give us an approximation of the process (6):\\
\[
\underline{m(\omega,a,t)} \Longleftrightarrow m(R^3 \backslash \omega, A \backslash a,t)
\]
\\
\\
$(1^o) \hspace{3cm}$ 
\begin{tabular}{|c|} \hline
\\
$\hspace{1cm}(a) \qquad m(D^\varepsilon \omega,A \backslash a,t) \Longleftrightarrow \underline{m(D^\varepsilon \omega,a,t)}$\\
\\
$(b) \qquad m(D^\varepsilon \omega,a,t) \Longleftrightarrow \underline{m(\omega,a,t)}$\\
\\ \hline
\end{tabular}\\
\\
The complex process $1^o$ gives us:
\[
m(D^\varepsilon \omega, A \backslash a,t) \Longleftrightarrow \underline{m(\omega,a,t)}.
\] 
 This process represents a part of the process (6). The second complex process is of the form:\\
\\
$(2^o) \hspace{3cm}$ 
\begin{tabular}{|c|} \hline
\\
$(a) \qquad m(d^\varepsilon \omega,a,t) \Longleftrightarrow \underline{m(d^\varepsilon \omega, A \backslash a,t)}$\\
\\
$\hspace{1cm}(b) \qquad m(d^\varepsilon \omega, A \backslash a,t) \Longleftrightarrow \underline{m(R^3 \backslash \omega, A \backslash a,t)}$\\
\\ \hline
\end{tabular}\\
\\
This complex process gives us
\[
m(d^\varepsilon \omega,a,t) \Longleftrightarrow \underline{m(R^3 \backslash \omega, A \backslash a,t)}.
\]
Hence the complex processes $(1^o), (2^o)$, represent together an approximation of the process\\
\[
\underline{m(\omega,a,t)} \Longleftrightarrow m(R^3 \backslash \omega, A \backslash a,t)
\]
let us analyse more precisely the processes $(1^o)$ and $(2^o)$.
\begin{center}
The process $1^o$.
\end{center}

The amount of the mass transported in the unit time in the process $(1^o)$, (a) to $m(D^\varepsilon \omega,a,t)$ is equal to
\[
\kappa^2 \int_{D^\varepsilon(\omega)}\int_a \int_{A\backslash a}\!\!\! M(x,t,\alpha,\beta)dx d\alpha d\beta
\]

Hence the density $\varrho(x,t)$ and the $\alpha$-density $\varrho(x,t,\alpha)$ of this mass are equal to
\[
\varrho(x,t) = \kappa^2 \int_a \int_{A \backslash a}\!\!\! M(x,t,\alpha,\beta)d\alpha d\beta,
\]
and
\[
\varrho(x,t,\alpha) = \kappa \int_{A \backslash a}\!\!\!M(x,t,\alpha,\beta)d\beta.
\]

The process $1^o$, (b) is of the type (5). Replacing in the description of the process (5) $\varrho(x,t,\alpha)$ by $\kappa \int_{A\backslash a}\!\!M(x,t,\alpha,\beta)d\beta$ and neglecting for simplicity the mass transport of diffusion type, we obtain that the amount of mass transported in the process $(1^o)$, (b) in a unit time to $m(\omega,a,t)$ is approximately equal to
\[
-\kappa^2\int_{\partial\omega}\!\!\!<n(x), \alpha >\biggl[\int_{A\backslash a}\!\! M(x,t,\alpha,\beta) d\beta \biggl]dx d\alpha =
\]
\[
= -\kappa^2\int_{\partial\omega}\int_a \int_{A\backslash a}\!\!\!<n(x),\alpha M(x,t,\alpha,\beta)>dx d\alpha d\beta.
\]
\begin{center}
The process $(2^o)$.
\end{center}

The density of mass transported in the unit time in the process $(2^o)$ (a) to $m(d^\varepsilon \omega, A\backslash a,t)$
is equal to
\[
\varrho(x,t) =\kappa^2\int_{A\backslash a} \int_a M(x,t,\alpha,\beta)d\alpha d\beta, \qquad x \in d^\varepsilon \omega,
\]
and the $\alpha$-density of this mass is equal to
\[
\varrho (x,t,\alpha)=\kappa \int_a M(x,t,\alpha,\beta)d\beta, \qquad x\in d^\varepsilon\omega, \alpha \in A\backslash a.
\]

The process $(2^o)$(b) is of the type (5), but in the opposite direction from $\omega$ to $R^3\backslash \omega$. Hence neglecting for simplicity the mass transport of diffusion type the amount of mass transported in the unit time in the process $(2^o)$(b) to the portion $m(R^3\backslash \omega,A\backslash a,t)$ is approximately equal to
\[
\kappa^2\int_{\partial\omega}\int_{A\backslash a}\!\!<n(x),\alpha>\biggl[\int_aM(x,t,\alpha,\beta)d\beta\biggl]dx d\alpha =
\]
\[
= \kappa^2\int_\omega \int_a \int_{A\backslash a}<n(x),\beta M(x,t,\alpha,\beta)>dx d\alpha d\beta.
\]

In the final description of the mass transport to $m(\omega,a,t)$ in the process (6) we have to take the last expression with  sign minus because it describes the mass learving $\omega$. Considering together the complex processes $(1^o)$ and $(2^o)$, we see that the amount of mass transported in the unit time in the process (6) to $m(\omega,a,t)$ is approximately equal to
\[
-\kappa^2\int_{\partial\omega} \int_a \int_{A\backslash a}<n(x),\alpha M(x,t,\alpha,\beta)+\beta M(x,t,\beta,\alpha)>dx d\alpha d\beta.
\]
Hence we obtain the following approximated expression for the boundary mixer:\\
\\
(20) $\hspace{3cm} B(x,t,\alpha,\beta) \approx \alpha M(x,t,\alpha,\beta)+\beta M(x,t,\beta,\alpha).$
\\

We have the integro differential system of 16 equations: (11) 2 equations (the mass conservation law), (14) 6 equations (the impulse conservation law), (18) 2 equations (the energy conservation law), (19) 6 equations (the impulsemomentum conservation law). The system contains 18 unknown functions, $\alpha$-quantities, and mixers:
\[
\varrho,\varepsilon,M,B_1,B_2,B_3,J_1,J_2,J_3,J^i_{B_j}; \quad i,j = 1,2,3.
\]
Basing on the relation (20), we introduce the following \underline{constitutive relation}:\\
\\
(21) $\hspace{2cm} B(x,t,\alpha,\beta)=\biggl[\alpha M(x,t,\alpha,\beta)+\beta M(x,t,\beta,\alpha)\biggl]b(x,t,\alpha,\beta),$
\\
where
\[
b: R^{10} \to R, \quad b = b(x,t,\alpha,\beta),
\]
is a new unknown function. In this way we reduce the number of the unknown functions to 16, and we obtain the general closed system of the model.

\vspace{0.5cm}
{\large{7. {\underline{Simplified Models.}}}}
\vspace{0.5cm}

The full integro-differential system of the model is very complicated and we are forced to seek for some simplified models. There exist many possibilities of such simplification. We shall give here the simplest of them. The main idea is to suggest some special form of the mass mixer $M(x,t,\alpha,\beta)$, for example\\
\\
(22) $\hspace{3cm} M(x,t,\alpha,\beta) = \Phi(\alpha,\beta)\mu(x,t,\alpha,\beta),$\\
\\
where $\Phi(\alpha,\beta)>0, \Phi(\alpha,\beta)=\Phi(\beta,\alpha)$, 
\[
\Phi(\alpha,\beta)=\cos \frac{\vartheta}{2}, \qquad \vartheta = \nless(\alpha,\beta),
\]
\[
\mu(x,t,\alpha,\beta) = \left\{
\begin{array}{l}
\varrho(x,t,\alpha)r(Dd)\;\mbox{for}\; d=|\beta|\varrho(x,t,\beta) - |\alpha|\varrho(x,t,\alpha) \geq 0,\\
\varrho(x,t,\beta) r(Dd)\;\mbox{for}\; d=|\beta|\varrho(x,t,\beta) - |\alpha|\varrho(x,t,\alpha) \leq 0
\end{array}
\right.
\]
$D= const >0$, and
\[
r(d) = \left\{
\begin{array}{lll}
-\frac{d}{1+d} & \mbox{for} & d \geq 0\\
-\frac{d}{1-d} & \mbox{for} & d \leq 0.
\end{array}
\right.
\]

A short justification of this form of mixer $M(x,t,\alpha,\beta)$ is as follows. Consider for $\alpha,\beta \in A$ two small fluid portions approximately moving with the velocities $\alpha$ and $\beta$. The mixer $M(x,t,\alpha,\beta)$ describes the amount of mass (positive or negative) transported in the unit time from $\beta$ to $\alpha$. Hence for the mixer (22), if $d=|\beta|\varrho(x,t,\beta)-|\alpha|\varrho(x,y,\alpha)\geq 0$ then the portion moving with velocity $\alpha$ loses (in the unit time), in favour of the portion moving with the velocity $\beta$, the amount of mass proportional to
\[
\cos \frac{\vartheta}{2}\varrho(x,t,\alpha)r(Dd).
\]
If $d=|(\beta)|\varrho(x,t,\beta)-|\alpha|\varrho(x,t,\alpha) \leq 0$, then the mass of a portion moving with the velocity $\beta$ loses (in the unit time), in favour of the portion moving with the velocity $\alpha$, the amount of mass proportional to
\[
\cos \frac{\vartheta}{2}\varrho(x,t,\beta)r(Dd).
\]

Our simplest model is based on the following assumptions.

$1^o$ We neglect the mixing process (6), $\underline{m(\omega ,a,t)} \Longleftrightarrow m(R^3\backslash \omega,A\backslash a,t)$, putting $B(x,t,\alpha,\beta)=0$.

$2^o$ We introduce the form (22) of the mass mixer $M(x,t,\alpha,\beta)$, defining in this way \underline{the mixing state} of the fluid. This is closely related to \underline{the impulse conservation} law, so as our mixer (22) decides how the impulse situation governs the mixing process.

$3^o$ In the simplified model, since $B(x,t,\alpha,\beta)=0$, the mass conservation law takes the following form\\
\\
(23) $\hspace{2cm}$
\begin{tabular}{|c|} \hline
\\
$\partial_t\varrho(x,t,\alpha)+<\alpha , grad_x\varrho>- E\Delta_x\varrho =\kappa\int_A M(x,t,\alpha,\beta)d\beta$\\
$\hspace{4cm} x \in R^3, \alpha \in A$\\
\\ \hline
\end{tabular}\\
\\
$4^o$ We use the energy conservation law in the form (17). The solution $\varrho(x,t,\alpha)$ of the closed system (23) and the thermodynamical state equation $p=f(\varepsilon,\varrho)$ allow to construct the Euler parameters $v(x,t), \varrho(x,t), \varepsilon(x,t), p(x,t)$.

Notice, that one may consider many different forms of mass mixers $M(x,t,\alpha,\beta)$. For example, we may introduce some thermodynamical or electrodynamical parameters to the definition of $M(x,t,\alpha,\beta)$ .

Observe that for the flow described by the $\alpha$-quantities the construction of the sets
\[
S(t,t^o,\omega)=\Phi_{t,t^o}(\omega)
\] 
must be based on the following relations:
\[
S(\tau + \Delta,\tau,\omega_\delta)\approx \bigcup_{y\in \omega_\delta} \biggl[y+\Delta {\bf{V}}(y,\tau)\biggl], \quad \tau \geq t^o, \omega_\delta = \{y:|y-x|<\delta\}; \mbox{hence}
\]
the smaller $\Delta>0$, the better approximation. The value of $\Delta >0$ should be chosen according to the properties of the sets ${\bf{V}}(x,t), t \geq 0$ that appear in the initial and boundary conditions. If ${\bf{V}}(x,t)$ changes in $x,t$ rapidly, then we have to take $\Delta > 0$ small enough otherwise $\Delta >0$ may be larger.

The question of determination of parameters $\varepsilon >0$ and $\kappa = \frac{3}{4\pi}(\frac{\Delta}{\varepsilon})^3$ is as follows. First we must realize that without strict definition of $\kappa$, the system (23) and the model are not defined in full. Taking $\kappa = const$ larger or smaller we decide that the mixing step in the flow considered will be larger or smaller, respectively.

For instance, this may be seen in our condition
\[
S(t+\Delta ,t,\omega_\varepsilon) \approx x+\Delta{\bf{V}}(x,t).
\]
The right side does not depend on $\varepsilon$. Hence, for big $\kappa$ and small $\varepsilon$ the set $\omega_\varepsilon$ is very small, and the very small mass portion $m(\omega_\varepsilon,t)$ must be spread on the set $x+\Delta{\bf{V}}(x,t)$ which does not depend on $\varepsilon$.

The first numerical experiments for our simplified model may be found in [2].
\\
\\
Bibliography.\\
\\
1. M. Burnat, K. Moszy\'{n}ski. "On some problems in mathematical modelling of turbulent flow"
 Journal of Technical Physics, 48, 3-4, 171-192, (2007). University of Warsaw, Institute
 of Applied Mathematics and Mechanics, preprint No. 161 (2007).\\
2. K. Moszy\'{n}ski. "Simplified non-Navier-Stokes model of turbulent flow
and its first numerical realization in 2D".
Submitted to preprints of the University of Warsaw, Institute of
Applied Mathematics and Mechanics (2011).\\
  
\end{document}